\documentclass[twocolumn,showpacs,preprintnumbers,amsmath,amssymb]{revtex4}

\usepackage[dvips]{graphicx}

\begin{document}

\title
{Ground-state energy and Wigner crystallization in thick 2D-electron
systems}

\author
{D. Jost 
}
\email[Email:\ ]{daniel.jost@ens-lyon.fr}
\affiliation{
Ecole Normale Sup\'erieure de Lyon, 46 All\'ee d'Italie 69364 Lyon, France}
\author
{M. W. C. Dharma-wardana}
\email[Email for correspondence:\ ]{chandre.dharma-wardana@nrc.ca}
\affiliation{
Institute of Microstructural Sciences,
National Research Council of Canada, Ottawa, Canada. K1A 0R6
}
\date{\today}
\begin{abstract}
The ground state energy of the 2-D Wigner crystal is
determined as a function of the thickness of the electron
layer and the crystal structure.
The method of evaluating the exchange-correlation energy is
tested using known results for the infinitely-thin 2D system.
Two methods, one based on the local-density approximation
(LDA), and another based on the constant-density approximation
(CDA) are established by comparing with quantum Monte-Carlo (QMC)
results. The LDA and CDA estimates for the Wigner transition
 of the perfect 2D fluid are at $r_s=38$ and 32 respectively,
compared with $r_s=35\pm5$ from QMC. For thick-2D layers as
found in
Hetero-junction-insulated-gate field-effect transistors,
the LDA and CDA predictions of the Wigner transition are at
$r_s=20.5$ and 15.5 respectively.
Impurity effects are
not considered here.
\end{abstract}
%
\maketitle
\section{Introduction}
Two-dimensional (2D) electron layers 
exist, for example, at the interface between GaAs and
Ga$_{1-x}$Al$_x$ As, or at the interface of a metal oxide and a
semiconductor. Such interfaces are important in field-effect
transistors and other devices.
The state of this 2D electrons system can be fluid or,
 at sufficiently low density, the electrons condense and
 crystallize\cite{ando,krav}.
We define
the 2D-density parameter $r_s$ given by $A/N=\pi{r_s}^2$ where $A$
 is the area
occupied by the $N$ electrons. The Wigner solid appears for $r_s\geq
35 a_0$ \cite{attac} where $a_0=\hbar^2/(m^*e^{*2})$ is the Bohr radius.
Here $m^*,e^*$ are effective parameters for the mass and the charge
of the electron, and absorb the dielectric constant, the band
mass and other material properties of the system. Thus
in GaAs, the effective atomic unit of energy is reduced from 27.21 eV
to the milli-Volt range. These (reduced) atomic units,
such that  $m^*=e^*=\hbar$=1 will be
assumed through out this paper.  
There has been many studies of 2D-electron liquids or Wigner crystals
\cite{attac, rapisarda,tanatceper,ceper,bonsall,iyakutti}, especially
 using quantum Monte Carlo (QMC)
simulations and other methods\cite{bonsall,prl2}
assuming that the 2-D layers are infinitely thin.
However, although the 2D electrons reside
in the $(x,y)$-plane, they have a transverse
 density $\eta(z)$ 
confined to the lowest subband of the hetero-structure \cite{ando}.
While the quasi-2D electron liquid has recently seen much
attention, both experimental\cite{tan, stormer} and
theoretical\cite{ssc,asgari,zhang,morsen}, the Wigner crystal in
thick 2D layers has not been followed up since the work of
Fujiki and Geldart
\cite{fuji2}. Fujiki et al., have
determined the effect of the 2D-layer thickness on the
electrostatic energy and found that the hexagonal lattice
is the most-stable crystal structure, as with the
 $\delta$-thin 2D layer ($\delta$2D).
However, they did not consider the effect of
exchange and correlation which is usually addressed
via Quantum Monte Carlo methods, or via a detailed analysis of
the correlated  phonons in the
electron crystal\cite{bonsall}.
We note that recent Hartree-Fock (HF) calculations of $\delta$2D
 Wigner crystals
using large plane-wave basis sets,
e.g, those of Trail et al\cite{trail}.,
seem to recover a HF energy nearly identical to the single-
Gaussian {\it harmonic} approximation for localized electrons.
Various aspects of such a model have been considered in
a brief but insightful paper by Nagy\cite{nagy,polini}. 
In this study we show that the single-Gaussian approximation,
and the local-density approximation (LDA),
can recover the QMC total
energy with surprising accuracy. We also show that a
method based on constructing a constant-density approximation
(CDA) to the inhomogeneous density\cite{ggsavin,ssc}
can be profitably used for calculating the
electrostatic potentials and the exchange-correlation 
energies of these systems.

The plan of the paper is as follows. In section~\ref{coul-sec} we
introduce the Hamiltonian, the effective Coulomb interaction
in quasi-2D layers, and calculate the electrostatic energy
of the lattice for several 2D-crystal structures. Here we use the CDA
to replace Fang-Howard type densities in the $z$-direction\cite{ando,ssc},
thus simplifying the analytical work.
The details of lattice-sum evaluations
are relegated to an appendix.
In section~\ref{perfect} we consider the 
$\delta$-thin 2D layer and present results for the
gaussian-localized model.
we also present the
exchange-correlation energy $E_{xc}$ calculation using the CDA and the
LDA.
The resulting total energy is very
close to that of QMC and recovers a liquid$\to$solid 
Wigner transition (WT) at
$r_s\sim$32 to 38, while the current QMC estimate is $r_s=35\pm5$. 
In section~\ref{gau-thick} we consider Gaussian-localized
2D systems with finite
thickness, for 2D layers found in HIGFETS.
That is, in systems where the layer thickness is 
also defined by the sheet density, as in Tan et al\cite{tan,stormer}.
 Here we have no QMC results for comparison. The total
energy of the quasi-2D Wigner crystal is compared with the
total energy of the quasi-2D liquid\cite{ssc}.
Here the WT is  found to occur at $r_s\sim$15 to 21
 in quasi-2D layers realized in clean HIGFETS,
\section{The Coulomb energies of 2D lattices}
\label{coul-sec}
The Hamiltonian of our system is, in atomic units,  
\begin{equation}
H=H_{ke}+H_{ee}+H_{eb}+H_{bb}
\end{equation}
where the first term is the kinetic energy of the electrons. 
The three remaining terms are the electron-electron interaction and
the interactions involving the uniform, static neutralizing background,
indicated by the subscript $b$.
This neutralizing background arises from a homogeneous
layer of donor ions which have acquired a positive charge after
donating their valence electrons in forming the 2D-electron layer.
We assume that the electron layer is confined
 near the plane $z=0$ and extends into
the region of $z>0$ due to the finite width of the envelope function.
 The donor ions are modeled by a homogeneous layer of
  positive charge of areal density $\rho_d=N/A$, situated at $z=-b_d$,
  where $b_d=|b_d|$ is a positive quantity.
The $z$-direction density is $\eta(z)$, and
 in the plane, an areal density $\rho_e(\mathbf{r})$ with $\mathbf{r}=(x,y)$.
 The subband distribution $\eta(z)$ is usually modeled by a
 Fang-Howard distribution $\eta_{fh}(z)=(1/2 b^3)z^2 e^{-\frac{z}{b}}$
 (n.b., our $b=1/b$ used in  Ref.~\cite{ando}),
 or various other forms, e.g, that of a quantum well. 
 The form of the density is obtained by fitting to a self-consistent
calculation of the Schrodinger equation for the electron motion in the
$z$-direction. In our work, we do not repeat this calculation, but simply
take the value of the parameter $b$, or other parameters
needed to define the self-consistent solution for the subband.
Moreover, as discussed below, such inhomogeneous
densities can be replaced by a constant-density slab
having an {\it equivalent electrostatic potential},
using the CDA discussed by Dharma-wardana\cite{ssc}. 
The CDA method \cite{ssc} involves
replacing an inhomogeneous density $\eta(z)$ by
a slab of constant-density
$\bar{\eta}$ of width $a$ linked to $\eta(z)$ by
\begin{equation}
\label{cda-eqn}
\bar{\eta}=\frac{1}{a}=\int {\eta(z)}^2 \,dz
\end{equation}
This equation has also been proposed by Gori-Giorgi
et al\cite{ggsavin}., in a method for
calculating system-adapted correlation
energies.
Using Eq.~\ref{cda-eqn} a Fang-Howard (FH)
 density of length scale $b$ can
 be replace by a homogeneous
density of width $a=(16b)/3$.
\begin{figure}[t]
\includegraphics*[width=8cm,height=6cm]{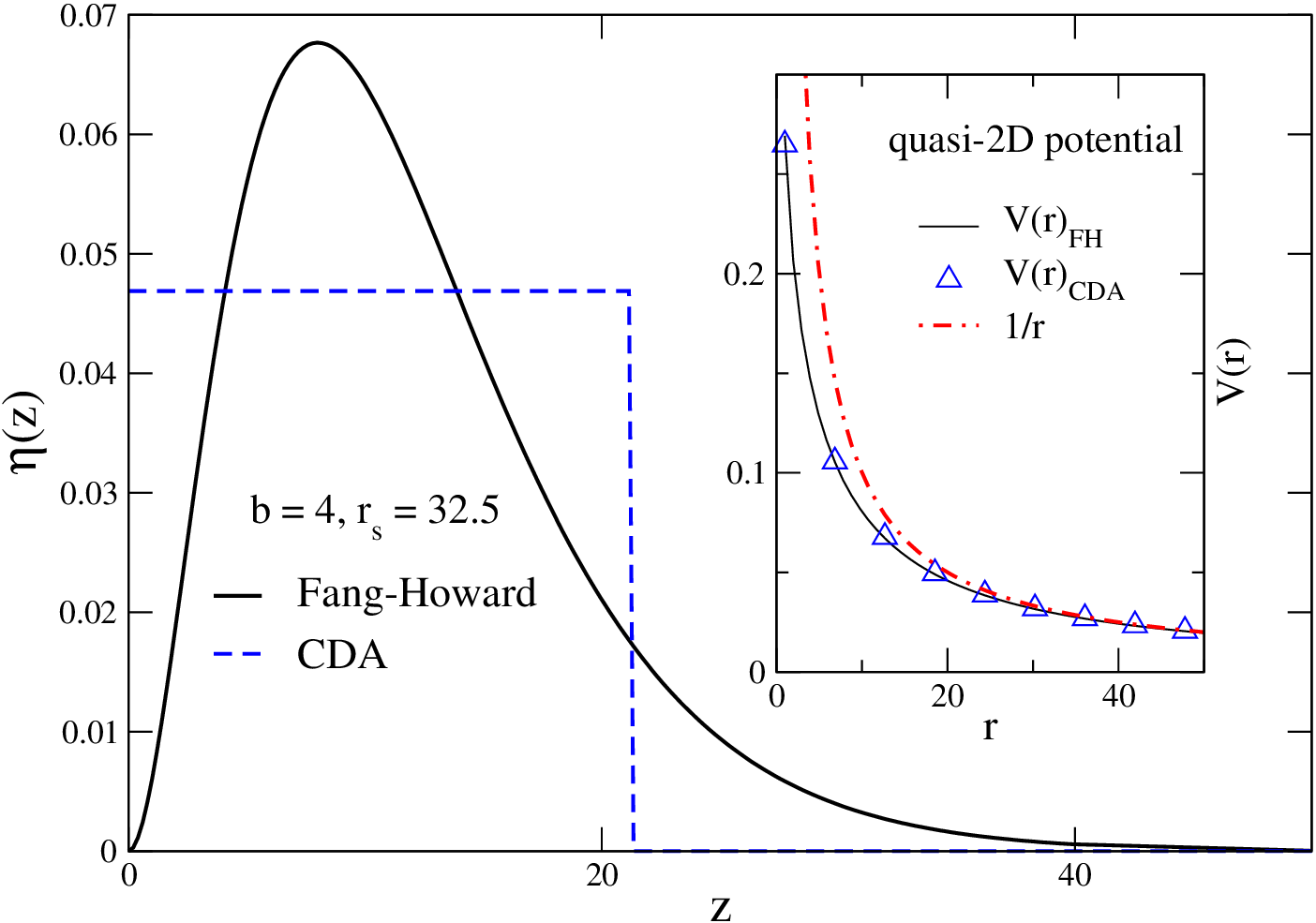}
\caption{(Color online) Profiles of the Fang-Howard density (solid curve)
 for $b=4$ and its 
  equivalent constant density approximation(CDA, dashed curve).
  Inset: the bare Coulomb potential $1/r$ and the Coulomb
  potential $F(r)/r$ modifed by the Fang-Howard profile. The
  triangles are calculated using the CDA. The CDA width
  a=21.33 for $b$=4 corresponds to a HIGFET at $r_s=32.496$.  
}\label{fang}
\end{figure}

Consider two electrons in a quasi-2D
layer separated by a distance $r$ in the 2D plane,
and located at $z_1$ and $z_2$, 
with a FH distribution $\eta(z)$ in the $z$ direction.
Then the Coulomb interaction is of the form
\begin{equation}
W(r)=\int d z_1\, d z_2
\frac{\eta(z_1)\eta(z_2)}{|r^2+(z_1-z_2)^2|^{1/2}}
\end{equation}

$W(r)$ may be written as $F(r)/r$ where
$F(r)$ is the form factor. No analytic form exists if 
$\eta(z)$ is the FH form, while the $q$-space form,
$F(q)2\pi/q$
is analytically available.
For GaAs/AlAs based HIGFET-like systems, it takes the form,
\begin{equation}F(q)=[1+\frac{9q}{8b'}+
\frac{3q^2}{8b'^2}][1+\frac{q}{b'}]^{-2}
\end{equation}
where $b'=1/b$ and follows the definition in
Ref.~\cite{ando}.
 However, if the FH distribution is
replaced by the CDA, then $F(r)$ and $F(q)$ are given by
\begin{eqnarray}
\label{cdmeqn}
W(r)&=&V(r)F(s),\;\; s=r/a, \;\; t=\surd{(1+s^2)} \\
F(s)&=&2s\left[\log\frac{1-t}{s}
\,\,+1-t\right]
\end{eqnarray}
and
\begin{eqnarray}
W(q)&=&V(q)F(p),\,\,\, p=qa, \,V(k)=2\pi/q\\
F(p)&=&(2/p)\{(e^{-p}-1)/p+1\}
\end{eqnarray}
The form factors $F(r)$, $F(q)$ are a
measure of the reduction of the strength of
the 2D interaction due to the thickness effect.
These results provide equivalent analytical
formulae for the FH-density,
and tend to the ideal 2D Coulomb
potential when the width $a$ tends to zero.
Also, for HIGFETS, it is known that the
FH-parameter $b$ is linked to the 2D density parameter
$r_s$. Hence it can be shown\cite{ssc} that
\begin{eqnarray}
\label{fh2w}
b&=&(2r_s^2/33)^{1/3}\\  
a&=&16b/3\\              
\beta&=&b/r_s=[2/(33r_s)]^{1/3}
\end{eqnarray}
Hence $\beta$, the FH parameter $b$
in units of $r_s$, and also the ratio $a/r_s$,
 i.e., ($z$-width)/(2D-disk radius)
decrease as $r_s^{1/3}$ with increasing $r_s$.
\subsection{Coulomb energy}
In the following we do not at first specify the form
of the transverse density $\eta(z)$.
In calculating the Coulomb energy $E_{Cou}$, i.e, the
electrostatic energy, we 
isolate the long-range contributions which cancel in the $q=0$
limit, since we are dealing with a homogeneous, neutralizing, static
background.  The total Coulomb energy
 is the sum,
\begin{equation}\label{ecoul}
E_{Cou}=\lim_{q \rightarrow 0} [E_{dd}(q)+E_{ee}(q)+2E_{ed}(q)]
\end{equation}
where
\begin{equation}
E_{dd}(q) =\frac{1}{2}\int d^2r \int
d^2r'{\rho_d}^2\,\frac{e^{i\mathbf{q}\cdot(\mathbf{r-r'})}}{|\mathbf{r-r'}|}
\end{equation}

\begin{widetext}
{\setlength\arraycolsep{2pt}
\begin{eqnarray}
E_{ee}(q) & = & \frac{1}{2} \int d^2r \int
d^2r'\rho_e(\mathbf{r})\rho_e(\mathbf{r'})e^{i\mathbf{q}
\cdot(\mathbf{r-r'})}\int_{0}^{\infty} dz \int_{0}^{\infty}
 dz' \frac{\eta(z)\eta(z')}{[(\mathbf{r-r'})^2+(z-z')^2]^{1/2}}\\
E_{ed}(q) & = & -\frac{1}{2}\int d^2r \int d^2r'\rho_d
\rho_e(\mathbf{r'})e^{i\mathbf{q}\cdot(\mathbf{r-r'})}
\int_{0}^{\infty} dz'\frac{\eta(z')}{[(\mathbf{r-r'})^2+(b_d+z')^2]^{1/2}}
\end{eqnarray}}
\end{widetext}
$E_{dd}$ is the interaction energy of the ions, $E_{ee}$ is 
the interaction of the electron layer and $E_{ed}$ is the
energy due to interaction between the ions and the electrons.
To calculate these terms, we proceed as in Fujiki
 and Geldart \cite{fuji2,fuji1}. 

\begin{equation}\label{edd}
E_{dd}(q)=\frac{\pi A{\rho_d}^2}{q}=\frac{N}{q{r_s}^2}
\end{equation}
We introduce the integral transformation
\begin{equation}\label{trans}
\frac{1}{|\mathbf{v}|}=\int_{0}^{\infty} \frac{dy}{\sqrt{\pi}}
 y^{-1/2}e^{-y|\mathbf{v}|^2}
\end{equation}
and note that
{\setlength\arraycolsep{2pt}
\begin{eqnarray}\label{eed}
E_{ed}(q) & = & -\frac{1}{2}\int d^2r\int d^2r'\rho_d
 \rho_e(\mathbf{r'})e^{i\mathbf{q}\cdot(\mathbf{r-r'})}
\int_{0}^{\infty} dz' \eta(z')\nonumber\\
& & \int_{0}^{\infty}\frac{dy}{\sqrt{\pi}}
 y^{-1/2}e^{-y|\mathbf{r-r'}|^2-y|b_d+z'|^2} \nonumber\\
& = &  -\frac{N}{2\sqrt{\pi}{r_s}^2}\int_{0}^{\infty}dy\, y^{-3/2}
e^{-\frac{q^2}{4y}}\nonumber\\
& &  \int_{0}^{\infty} dz'\eta(z')e^{-y|z'+b_d|^2}
\end{eqnarray}}
For $E_{ee}$, we use a lattice sum technique based on the
 $\theta$ Jacobi function, Eq.(\ref{theta1}),
  and its imaginary transform, Eq.(\ref{theta2}), given below:
\begin{equation}\label{theta1}
\theta(z,X)\equiv \sum_{l=-\infty}^{\infty} e^{2\pi lz}e^{-\pi l^2 X}
\end{equation}
\begin{equation}\label{theta2}
\theta(z,X)=\frac{e^{\pi z^2/X}}{\sqrt{X}}\theta(\frac{z}{iX},\frac{1}{X})
\end{equation}
We decompose the lattice into rectangular sublattices
 indicated with sublattice vectors $\mathbf{\rho_j}$. 
 So, the position vectors of the electrons on
  nodes $I$ and $J$ are given by
\begin{center}
$\mathbf{r_I}=m a_1 \hat{x} + n a_2 \hat{y},
 \quad \mathbf{r_J}=(m'a_1+\rho_{j}^{x})
 \hat{x}+(n'a_2+\rho_{j}^{y})\hat{y}$
\end{center}
where $m,m',n,n'$ are integers,
 $a_1,a_2$ are lattice constants of sublattices.
For example, in a square lattice $a_1=a_2$ and
 $\{ \mathbf{\rho_j} \}=\{(0;0)\}$,
 in a hexagonal lattice $a_2=\sqrt{3}a_1$ and $ \{ \mathbf{\rho_j}
 \}=\left\{(0;0),\left ( \frac{a_1}{2};\frac{a_1\sqrt{3}}{2} \right)\right\}$.
To proceed further, we need to specify the form of the density.
 If the electrons are assumed to be
exactly localized on the nodes of the crystal, then
\begin{equation}\label{rhoefg}\rho_{e\delta}(\mathbf{r})=\sum_{I}
  \delta(\mathbf{r-r_I}).\end{equation}
Such exact localization of the electrons provides the model for
the classical electrostatic energy, i.e, {\it the
Madelung energy}. 
In the quantum calculation
 we suppose that each electron is localized around a node $I$ of the
lattice and the wavefunction  is
taken to be a gaussian normalized over the 2D plane,
\begin{equation}\label{phi} \phi_I(\mathbf{r})=\sqrt{\frac{2\alpha}{\pi}}e^{-\alpha
    (\mathbf{r-r_I})^2}\end{equation}
The parameter $\alpha$ is chosen to  
minimize the total energy. Hence the localized density is
\begin{equation}\label{rhoegauss}
  \rho_{eG}(\mathbf{r})=\frac{2\alpha}{\pi}\sum_{I} e^{-2\alpha
    (\mathbf{r-r_I})^2}
\end{equation}
The gaussian-width parameter $\alpha$ is of the form
$a/r_s^{3/2}$, with $a$ taking a lower-bound value of 0.5 (see
Ref~\cite{nagy}).
These two forms of the density will be studied below, and the Gaussian
approximation will be justified by comparison with
results from detailed plane-wave calculations.
\subsection{Calculation with the $\delta$-distribution}
\label{gau-thin}
Using Eq.(\ref{trans}) and (\ref{rhoefg}) we have
\begin{eqnarray}
\label{delta-ee}
E_{ee}(q)&=&\int_{0}^{\infty} \frac{dy}{2\sqrt{\pi y}}f(y)
 \sum_{I\ne J}e^{i\mathbf{q\cdot(r_I-r_J)}}e^{-y|\mathbf{r_I-r_J}|^2} \\
f(y)&=&\int_{0}^{\infty} dz \int_{0}^{\infty} dz'\eta(z)\eta(z')e^{-y(z-z')^2}
\nonumber
\end{eqnarray}
The details of the evaluation are given in the Appendix.

\begin{table}
\caption{The 
Madelung energy, $E_{Cou}$ per electron are given for
  different values of the Fang-Howard
   parameter $\beta=b/r_s$ for hexagonal(hex), square(sq),
  rectangular(rec), centered rectangular(cr) lattices defined by
  their unit vectors $a_1$:$a_2$. The $r_s$ parameter
  in the corresponding HIGFET, Eq.~\ref{fh2w},
  is also given. 
 The energies are in 
   units of $1/r_s$. Thus the Madelung energy in Hartrees
    for a $\delta$-thin
   hexagonal lattice is $-1.106103/r_s$}
\label{tabecoul}
\begin{ruledtabular}
\begin{tabular}{lccccc}
HIGFET $r_s$                         &$\infty$ & 60606      & 60.606      & 0.06060\\
$\;\;\;(a1:a2)$ $\beta\to$\rule{0pt}{2.6ex} & $0$     &  $10^{-2}$ & $10^{-1}$ & $1$ \\

\hline 
hex($\sqrt{3}:1$) \rule{0pt}{2.6ex} & $-1.106103$ &  $-1.052959$ & $-0.591433$
& $3.144793$ \\

cr($\sqrt{2}:1$) \rule{0pt}{2.6ex} & $-1.104080$ &  $-1.050937$ & $-0.589507$ &
$3.145401$ \\

sq(1:1) \rule{0pt}{2.6ex} & $ -1.100244$ &  $-1.047103$ & $-0.585854$ & $3.146555$ \\

rec($\sqrt{2}$:1) \rule{0pt}{2.6ex} & $-1.078201$ &  $-1.025072$ & $-0.564890$ &
$3.153217$\\

rec($\sqrt{3}:1$) \rule{0pt}{2.6ex} & $-1.042843$ &  $-0.989733$ & $-0.531301$ &
$3.163948$ \\

\end{tabular}
\end{ruledtabular}
\end{table}

 We have evaluated $E_{Cou}$, Eq.~\ref{ecoul} for
 different lattices:
  square, rectangular, hexagonal and
  centered-rectangular. The Coulomb energy depends
  only on $\beta=(b/r_s)=(3a)/(16 r_s)$, $r=(a_2/a_1)$
   and $\{\mathbf{\rho_j}\}$. 
Our numerical calculations of $E_{Cou}$ are summarized in
  Table \ref{tabecoul}. Results for $\beta=10^{-2}$ are at
 unrealistically low HIGFET densities, but are of formal
 interest. Results for even smaller values
 of $\beta$ may be found in Fujiki et al\cite{fuji1,fuji2}.
 A comparison with the results of Ref.~\cite{fuji2}
  shows that our results are in
  agreement when a geometrical term arising from the slight
  difference in the models is taken into account.
   (As seen from the details given in the appendix, 
    we have an additive term 
  $N(2a/3)/{r_s}^2=N(32b/9)/{r_s}^2$ in our calculation while Fujiki
  and Geldart have $N(33b/8)/{r_s}^2$. Agreement is obtained if
  we replace our term by theirs).

   It is seen that the total Coulomb energy increases as
  $\beta=b/r_s$ increases for all cases studied. The hexagonal lattice
  has the lowest energy for all $\beta$. Moreover, there is
  no crossing between the different energy curves for any of the
  lattice structures.

The dependence of the total Coulomb energy of the
 centered-rectangular lattice
and rectangular lattice as a function of the
 ratio $r=a_1/a_2$ for the quasi-2D system
remains similar to the $\delta$-thin case. Two equivalent minima
at $r=\sqrt{3}$ and $1/\sqrt{3}$ correspond to
 the hexagonal structure. For the
rectangular lattice, the minimum corresponds to $r=1$, i.e., to the
square structure.  We choose the range $\beta$=0.05 to 0.5, which
corresponds to $r_s\sim$ 0.5 to $\sim$500
and fit the Madelung energy of the stable hexagonal
lattice (see Table \ref{tabecoul}) to the analytic form
\begin{equation}
\label{thick-madelung}
E_{Cou}(r_s,\beta)=\sum_{i=0}^{i=4}c_i\beta^i/r_s
\end{equation}
where $c_0=-1.106103$, $c_1$=5.34722, $c_2=-2.15257$, $c_3$=1.48663, and
$c_4=-0.430473$. 

\begin{figure}
\includegraphics*[width=8cm,height=6cm]{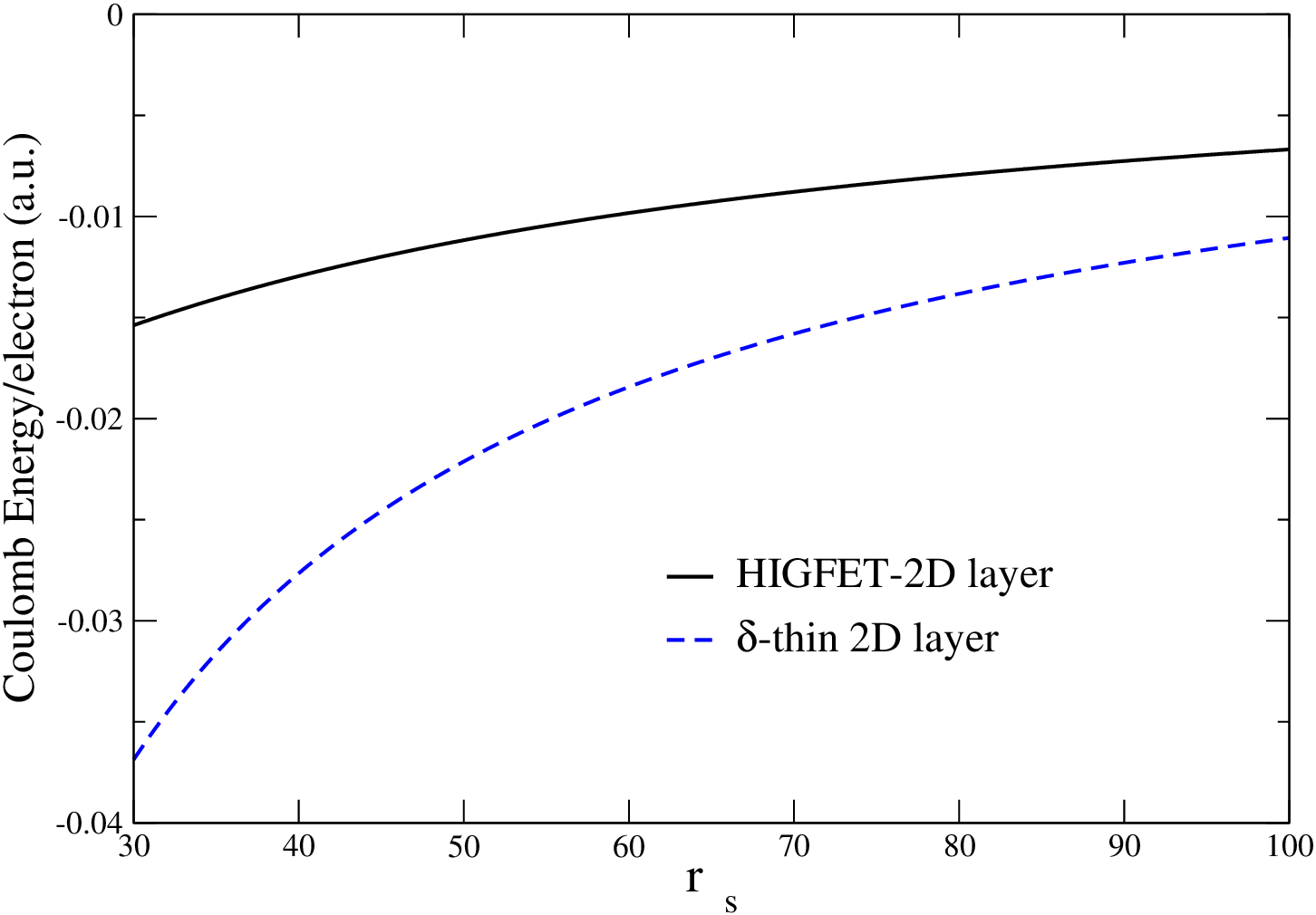}
\caption{(Color online) (Coulomb energy per electron in atomic units for a perfectly 2D
system  and for 2D layers in a HIGFET, using
Eq.(\ref{fh2w}) to define the thickness.}\label{bfh}
\end{figure}

In Figure \ref{bfh}, we have plotted the Coulomb 
energy as a function of $r_s$
using Eq.(\ref{fh2w}) to relate the thickness to the $r_s$ value.
We observe that the thickness of the system
has a significant effect on the energy.
\subsection{Classical calculation with the gaussian distribution.}
If the electron distribution at each site were
a gaussian, the classical {\it electrostatic} 
energy, $E_{ee}$, can be calculated using the same
techniques as before (Appendix).
{\setlength\arraycolsep{2pt}
\begin{eqnarray}
E_{ee}(q,\alpha) & = & \frac{1}{2\sqrt{\pi}}\int_{0}^{\infty}
dy\,y^{-1/2}\left(\frac{\alpha}{y+\alpha}\right)e^{-\frac{q^2}{4(y+\alpha)}}\nonumber\\
& & \sum_{I\ne
  J}
e^{-\frac{y\alpha}{y+\alpha}(\mathbf{r_I-r_J})^2}
e^{-i\frac{\alpha}{y+\alpha}\mathbf{q\cdot(r_I-r_J)}}
\end{eqnarray}}
We use the same integral separation with $E_{ee}^{<}$ and $E_{ee}^{>}$,
the Jacobi function $\theta$ and its transformation.
We may verify that when $\alpha$ tends to zero, that
is to say the gaussian distribution tends to the $\delta$-distribution,
$E_{ee}$ reduces to the Madelung energy of the
 previous section. Also, if there is no effective
 overlap among the gaussian distributions,
 the distributions can be replaced by equivalent 
 point charges at the lattice sites and the Coulomb
 energy should reduce to the Madelung
 energy. However, as already remarked by
 previous authors\cite{nagy,polini}, the charge 
 is not perfectly contained within the Wigner-Seitz
 disk in the 2D problem.
  The variations of the thickness and of the lattice type give results
 similar to the $\delta$-thin case. 
 We consider the variation of $\alpha$ to minimize the
 total energy within a quantum calculation, and hence do not develop this
 classical calculation any further. 
\section{Perfectly two-dimensional systems}\label{perfect}
 In this section we consider a perfect,
  i.e., $\delta$-thin 2D layer within a
  Kohn-Sham density-functional approach\cite{kohn}
  to the quantum mechanics of the problem. Since the 
  $\delta$-thin 2D system has been studied extensively, we use it
  as a reference system to examine the LDA and the CDA as
  useful tools for calculations of $E_{xc}$ of Wigner
  crystals. The Hohenberg-Kohn theorem
  asserts that the total energy is a functional
   of the one-electron density,
  and that it is a minimum for the true density distribution. We model the
  one-electron density as a sum of gaussians centered on each lattice
  site, and hence the variational problem reduces to a determination
  of the width parameter $\alpha$ of the Gaussian that minimize the
  total energy. 
  The total energy of the system at a given $r_s$ can be written as:
  \begin{equation}
  \label{dft-etot}
  E_T= E_{HF}(\alpha,r_s) +E_{xc}(\alpha,r_s)
  \end{equation}   
  where $E_{HF}(\alpha,r_s)$ is the Hartree-Fock energy of an electron.
  It will be seen that this is effectively
  the energy of an electron on a single site,
   and moving in
  the potential well created by the gaussian
  distributions on other sites. If the gaussians were
  perfectly localized, the Coulomb energy would not
  depend on $\alpha$.  The effect of the overlap
  can be easily included in the variational problem,
  with the energy given by $<\psi|H|\psi>/<\psi|\psi>$, and
  this has an effect for small $r_s$. Here $\psi$ is 
  a Slater determinant of gaussians. For the hexagonal lattice,
  the 
  overlap contribution from two nearest-neighbour gaussians is
  $$s_{ij}(r_s)=\exp[-(\alpha/2)(1.09r_s)^2]$$
  where $1.09r_s$ is the nearest-neighbour distance. 
  Unless  the contrary is stated, the results reported here
  will include the overlap correction.
  The $\alpha$ which minimizes
  the Hartree-Fock problem is not the same as that which minimized
  the total energy inclusive of $E_{xc}$. In the next section we
  look at the problem without $E_{xc}$

\subsection{The Hartree-Fock energy $E_{HF}$}
\label{hf-sec}
The Hartree-Fock energy is composed of the classical Madelung energy
which defines a constant energy term, plus the quantum mechanical
energy associated with the motion of the electron in the field of the
other electrons. Since the electrons are strongly
localized, especially for large $r_s$, 
a Slater determinant made up of one gaussian function
at each lattice site is commonly assumed. 
The total energy consists of a
kinetic energy  term and a potential energy term. These two
terms are equal by the virial theorem and hence we only
need to evaluate the kinetic energy. 
Usually, Hartree-Fock energies contain a sizable exchange
contribution. However, the localized-gaussian exchange energy is
easily shown to be negligible,
 and we called it the Wigner-exchange
energy, $E_{Xwc}$.  

In Table~\ref{hf-tab} we compare our localized-gaussian 
(harmonic) calculation with the results of the extensive
plane-wave HF calculation by Trail et al\cite{thank-trail}.
\begin{table}
\caption{Comparison of the plane-wave calculation\cite{trail}
 of the HF energies $E_{HF}$ of the $\delta$-thin 2D hexagonal Wigner
lattice with the single-Gaussian harmonic lattice energies.
$E^*_{har}$ and $E_{har}$ are energies without and
with the overlap corrections.
}
\begin{ruledtabular}
\begin{tabular}{ccccccc}
$r_s\to$ & 20& 30&40 & 60 \\
\hline
-$E_{HF}\times10 $   &0.447270 & 0.311642 &0.239528 &0.164036\\ 
-$E^*_{har}\times10$ & 0.447155 & 0.311786 &0.239822 & 0.164530\\
-$E_{har}\times10$   & 0.437058 & 0.308344 &0.238326 & 0.164113\\
\end{tabular}
\end{ruledtabular}
\label{hf-tab}
\end{table}
The results shown in Table~\ref{hf-tab} 
show that the localized single-gaussian model  is adequate to
describe the Hartree-Fock approximation
for this system\cite{trail2}.

Note that our calculation is effectively an ``Einstein model'' of
oscillators, and the kinetic energy is
given by,
\begin{equation}
E_K(\alpha)= -\frac{N}{2}<\phi_I|\nabla_I^2|\phi_I>=N\alpha
\end{equation}
The 
gaussian width which minimizes the energy
may be fitted by the form  $\alpha=0.6263/r_s^{1.57}$.

Since the exchange of electrons actually involves a delocalization
process, we believe that the exchange integral evaluated with fixed
gaussians does not lead to a true evaluation of the exchange in
these systems.
The Wigner-exchange energy between
 two electrons of spin $s_i,\,s_j$, is by definition
{\setlength\arraycolsep{2pt}
\begin{eqnarray}\label{excdef}
E^{ij}_{Xwc}& = & -\int d^2 r_i\,d^2 r_j
\phi_I(\mathbf{r_i})\phi_J(\mathbf{r_j})\frac{1}{r_{ij}}
\phi_I(\mathbf{r_j})\phi_J(\mathbf{r_i})\nonumber\\
& = & -\sqrt{\alpha\pi}e^{-\alpha(\mathbf{r_I-r_J})^2} \delta_{s_i,s_j}
\end{eqnarray}}
We can define a polarization parameter
$\zeta=(N_{\uparrow}-N_{\downarrow})/N$, therefore
\begin{equation}
\label{wigner-X}
E_{Xwc}(\alpha,\zeta)=-\frac{1}{2}\sum_{i\ne j} E^{ij}_{Xwc}
\end{equation}
This 
$E_{Xwc}$, may be safely neglected for the values of $\alpha$
occurring in this problem. 
\subsection{The CDA exchange and correlation energies}
\label{correlperf}
The correlation energy is the most difficult
object to calculate, and QMC has been the
preferred approach, even though this requires a
 a major numerical effort.
However, for an uniform density profile,
 the 
correlation energy is well known \cite{attac}. Hence, as in 
Eq.i~\ref{cda-eqn}, we map the inhomogeneous density in the
$(x,y)$ plane, $\rho(\mathbf{r})$, to a
homogeneous form via the $<\rho(\mathbf{r})^2>$ average of the CDA method.
Given a gaussian distribution,
{\setlength\arraycolsep{2pt}
\begin{eqnarray}
\bar{\rho} & = & \frac{1}{\pi {r_s}^2}\int d^2 r\, |\phi(\mathbf{r})|^2
|\phi(\mathbf{r})|^2 \nonumber \\
& = & \frac{\alpha}{\pi^2 {r_s}^2}
\end{eqnarray}}
We define the effective $r_s$ parameter  $\bar{r_s}$
corresponding to the CDA density by $\bar{\rho}=1/(\pi{\bar{r}_s}^2)$,
\begin{equation}
\label{rsCDA}
\bar{r}_s=r_s\sqrt{\frac{\pi}{\alpha}}
\end{equation}
The correlation energy in the CDA, $E_{c}^{cda}$ for the
inhomogeneous distribution, inclusive of spin-polarization effects,
is now evaluated using $\bar{r}_s$ in any of the
well-known 2D-functionals\cite{attac}.
Note that for typical values of $\alpha$ at $r_s=20$, the
CDA density parameter is $\sim 400$, while at $r_s$=100,
it becomes $\sim$7000. Thus
we see that the CDA replaces the inhomogeneous fluid with
sharp Gaussian peaks by a uniform, {\it ultra-low-density}
 2D fluid.
In calculating $E_c(\bar{r}_s)$ using, say, the formula
due to Attaccalite et al., a difficulty arises since it is
fitted to a maximum $r_s$ of 40, together with
asymptotic forms, while
the CDA calls for $r_s$ values which are one or two orders
bigger. Nevertheless, we find
surprisingly good results (see below).

At this point we ask if the exchange energy, evaluated
for this ultra-low density fluid, should also be
included. We believe that this is indeed the case. The fixed-gaussian
Wigner-exchange, Eq.~\ref{wigner-X} simply does not allow
any exchange, and ignores the possibilities of tunneling, 
ring-exchange etc., that exist in the system. We consider that the
estimate of exchange obtained from the ultra-low density
fluid of the CDA accounts for such exchange effects. This
point of view is justified {\it post-facto} by the good
agreement of our total energies with the QMC total energies.

\subsection{The LDA exchange and correlation energies}
Another approach to replacing the inhomogeneous electron density
by a homogeneous fluid-density is the 
local-density approximation (LDA) \cite{kohn}. 
Here a uniform density corresponding to each
local density $n(r)$ is invoked.
 Thus a local-density parameter
$\bar{r}_s(r)$ is defined by 
\begin{equation}
\label{rsLDA}
\frac{1}{\pi\bar{r_s}^2}=
\frac{\rho(r)}{\pi{r_s}^2}\quad \Longrightarrow \quad
\bar{r}_s(r)=r_s e^{\alpha r^2} \sqrt{\frac{\pi}{2\alpha}}.
\end{equation}
Hence, knowing the exchange-correlation
energy density $e_{xc}$ for a homogeneous
system, the exchange-correlation energy
 of the inhomogeneous system is
given by
\begin{equation}\label{xclda}
E_{xc}^{LDA}=\int d^2r \, e_{xc}[\bar{r}_s(r)]\,\rho(r)
\end{equation}
Just as in the CDA, the LDA demands the evaluation
of $E_{c}$ at densities which are beyond
the range of the standard fits. Thus LDA needs
$r_s(r)\sim 300$ to 5000 at $r_s=20$, i.e., a
little less extreme than the CDA.
Hence, some of the shortcomings of the LDA may
also be due to poorly known correlation energies at 
the exceptionally high $r_s$ values that are
required.
The LDA can be further improved by including
gradient corrections. However, we have not 
included them in this study.  
\subsection{Minimization of the total energy $E_T$}
\label{perfmin}
We have now all the energy contributions
needed to calculate the ground-state energy of a
perfect two-dimensional
 (i.e., $\delta$-thin) Wigner crystal at a given value of
 the density parameter
$r_s$.  
The energy minimum with respect to $\alpha$
 is found to be insensitive to the polarization of
the lattice. This in agreement with previous studies
\cite{attac,rapisarda,tanatceper,zhu}. 
\begin{table}
\caption{Results of energy minimization for a hexagonal
lattice. The Madelung energy $E_M=-1.106103/r_s$
has been subtracted out from the total energy.
The CDA and LDA results are compared with the  GFMC
 calculations of Tanatar
 \& Ceperley, T-C \cite{tanatceper}
 The energies are in $10^{-2}$ atomic
units. }
\label{tabpol}
\begin{ruledtabular}
\begin{tabular}{ccccccc}
$\;\;\;r_s\to$  & 20   & 30    & 40   & 50    & 60 \\
\hline
CDA  &0.9247 & 0.4824 &0.3059 & 0.2156 &0.1625 \\

LDA  &0.9404 & 0.4899 &0.3102 & 0.2185 &0.1645 \\

T-C  &0.9167 & 0.4983 &0.3234 & 0.2313 &0.1758 \\

\end{tabular}
\end{ruledtabular}
\end{table}
In Table \ref{tabpol}, we give the energy correction to the
Madelung energy
obtained by the minimization of $E_T$,  using the CDA or the LDA 
for evaluating the exchange-correlation effects,
together with the results of previous work \cite{tanatceper}.
QMC results by Rapisarda and Senatore\cite{rapisarda} are
very similar to those of Tanatar et al., and the agreement
is similar.  
The optimal $\alpha$ which minimizes the energy
is found to be given by
$\alpha=a/r_s^{3/2}$
with a=0.639 for both CDA and LDA approaches. Here
we note that Nagy's method\cite{nagy} predicts
a value $\alpha_{ng}=0.5/r_s^{3/2}$ as a lower bound.
A crucial test of the accuracy of the CDA and LDA would lie
in their ability to predict the liquid$\to$solid phase transition.
This is addressed in section~\ref{wpt-sec}.
The total energy can be represented by:
\begin{equation}
\label{etot-fit}
E_T(r_s)=\frac{a_1}{r_s}+\frac{a_2}{{r_s}^{3/2}}+\frac{a_3}{{r_s}^2}+O({r_s}^{-5/2})\quad
r_s\gg 1
\end{equation}
where $a_1=-1.106103$ is the Madelung constant and $a_2$ is the
zero-point energy of the lattice. 
We determined the coefficient $a_3$ by a least-square fit. 
The results are summarized in Table~\ref{tab-123},
together with previous results.
\begin{table}
\caption{Coefficients $a_1-a_3$  in Eq.~\ref{etot-fit}
 fitting the CDA and LDA
total energy (for the range $r_s$=20 to 100) are compared
with previous work. 
}
\label{tab-123}
\begin{ruledtabular}
\begin{tabular}{ccccccc}
  &CDA & LDA &BM\cite{bonsall}  & RS\cite{rapisarda} & TC\cite{tanatceper} \\
\hline
$-a_1$ &  1.1061  & 1.10610   & 1.1060  & 1.104715 &1.10610 \\ 
$a_2$  &  0.8142  & 0.8142     & 0.8142  & 0.7947  &0.8142  \\
$a_3$   & 0.2456  & 0.1194     & ...     & 0.07338  &0.0254 \\
\end{tabular}
\end{ruledtabular}
\end{table}
 These results justify
our use of the CDA and the LDA for evaluating the
total energy of quasi-2D Wigner crystal phases for which there
are no QMC calculations as yet.

\section{Influence of the thickness}
\label{gau-thick}
 We consider a
quasi-2D electron crystal where each electron is localized
at each lattice site with a gaussian distribution centered on each site
in the $(x,y)$-plane,
while the $z$-extension may typically have the form of a Fang-Howard density.
As before, such $z$-distributions can be replaced by a constant-density
form for ease of calculations. Also, we assume that the 2D layers
are in HIGFETS, and as such the FH-parameter $b$ is automatically
specified (via Eq.~\ref{fh2w} )when the $r_s$ parameter defining
the 2D-layer density is specified.

The kinetic energy and the harmonic energy of the quasi-2D
system are still given by $E_K(\alpha)=N\alpha$ since
this is a result of the assumed gaussian form of the wavefunction.
However, the simple Coulomb potential $1/r$ has changed to
 $F(r)/r$ where $F(r)$ is the form factor arising from the
subband distribution.  The Wigner-exchange energy, i.e.,
the exchange between two localized electrons is now even
weaker than in Eq.~\ref{excdef}. Hence this type of
exchange is totally negligible.

\subsection{The evaluation of $E_{xc}$ for thick-2D layers using
 CDA and LDA.}
 As  described in Eq.~\ref{fh2w}, the z-distribution
is mapped onto a uniform slab of width $a$; in HIGFETS
this is
directly related to the $r_s$ parameter in the 2-D plane.
The inhomogeneous 2-D distribution in the plane is also
mapped onto a homogeneous distribution via the CDA 
as in Eqs.~\ref{rsCDA}
or as in Eq.~\ref{rsLDA} for the LDA.
Both CDA and LDA require a knowledge of the $E_{xc}(r_s,a)$
for quasi-2D uniform systems with a layer width $a$. Parametrized forms
of the exchange energy and the correlation energy
of uniform quasi-2D electron fluids have been given by
Dharma-wardana\cite{ssc}. The exchange energy $E_x(r_s,a)$
is given in the form $E_x(r_s,\zeta)Q(r_s,a,\zeta)$ where
$E_x(r_s,\zeta)$ is the well-known exchange energy of the
$\delta$-thin system, while $Q(r_s,a,\zeta)$ is a form factor.
The correlation energy of quasi-2D layers in HIGFETS
 is given in Ref.~\cite{ssc}
as an interpolation involving a form for electron ``rods'' interacting
via a logarithmic potential (as is the 
case for small $r_s$), and for 3D-like electrons when $r_s$,
and the thickness $a$
become large. The Wigner crystallization regime involves
the latter regime. For details of these parametrizations,
the reader is referred to Ref.~\cite{ssc}. Since the WT involves
small energy differences, we have actually done an explicit
calculation instead of using the fits.

\begin{figure}
\includegraphics*[width=8cm,height=6cm]{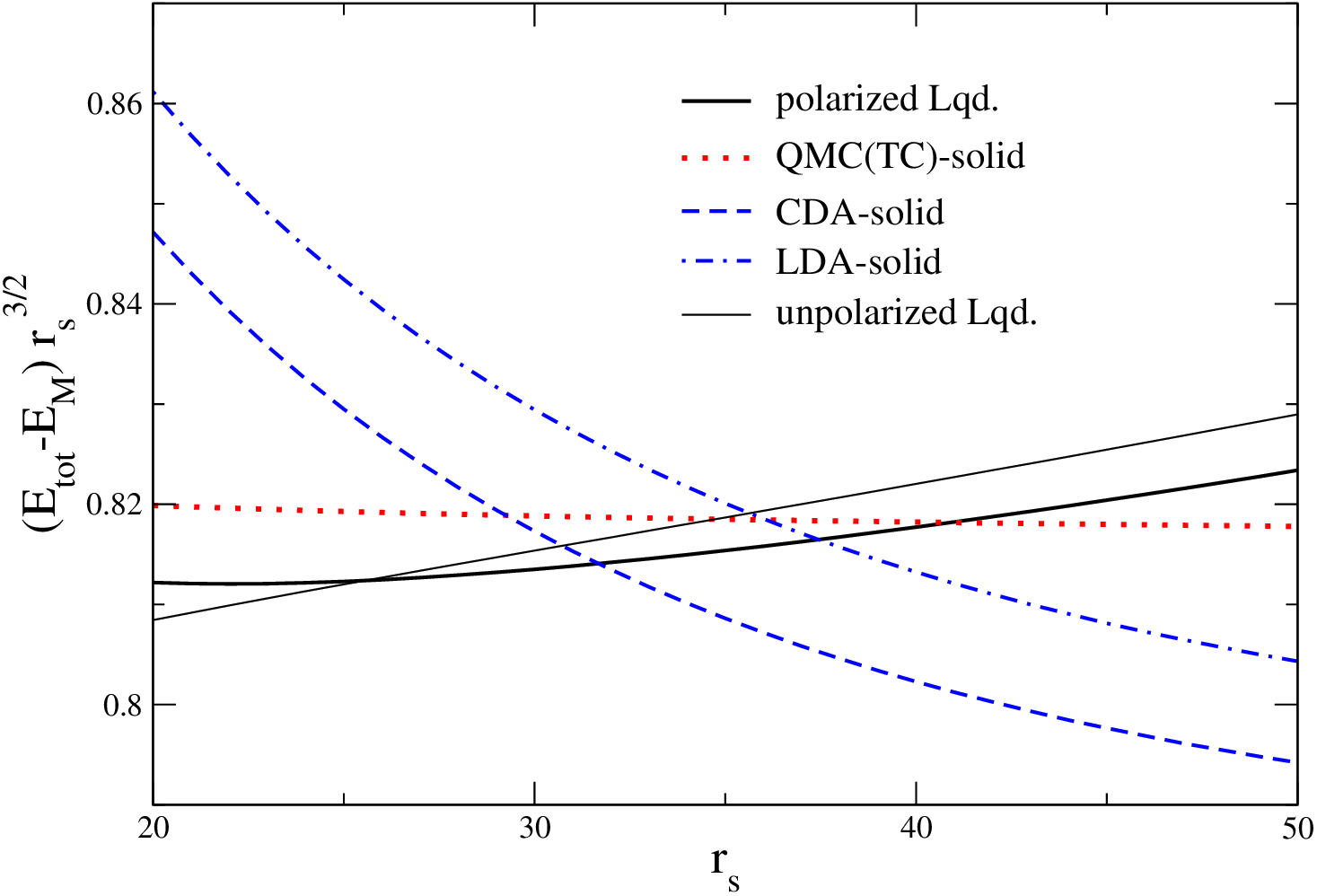}
\caption{(Color online) Comparison of liquid and solid-phase energies.
$(E-E_M){r_s}^{3/2}$ where
$E_M=-1.106103/r_s$ and $E$ is the unpolarized or
fully-polarized fluid energy, or the solid energy
$E_{cda}$,  $E_{lda}$ or from QMC.}
\label{figwpt}
\end{figure}

\subsection{Minimization of the total energy $E_T$}

As in Sec.\ref{perfmin}, we minimize the total energy as a function of
$\alpha$ for a given $r_s$. Here, the
energy is more sensitive to the spin-polarization
$\zeta$  than in the perfect crystal even
if the difference is very small. The unpolarized crystal is more stable
than the polarized one. So we present results for the unpolarized
system.
The values of $\alpha$ which minimizes the energy can also be fitted by the
same form as in Sec.\ref{perfmin}. We obtain
\begin{equation}
\alpha_{cda}=\frac{0.619}{{r_s}^{3/2}} \quad\mathrm{and}\quad
\alpha_{lda}=\frac{0.627}{{r_s}^{3/2}}
\end{equation}
We have also fitted the total energy. Here the Madelung energy is 
 the  $E_{Cou}$ given in Eq.(\ref{thick-madelung}) and we use the
usual expansion in inverse $r_s^{3/2}$ etc.
\begin{eqnarray}
E_{T}^{cda} & = &
E_{Cou}(r_s)+\frac{0.68597}{{r_s}^{3/2}}+\frac{0.321652}{{r_s}^2} \\
E_{T}^{lda} & = &
E_{Cou}(r_s)+\frac{0.708977}{{r_s}^{3/2}}+\frac{0.357242}{{r_s}^2}
\end{eqnarray}
We remark that the total energy has a minimum as a function of $r_s$. This
minimum is located around $r_s\sim 26$ and its value is $\sim -0.011$ a.u.
A comparison of  liquid phase and Wigner crystal in HIGFETS is given in
Table~\ref{tabpol2}. These total energies include the $2a/3r_s^2$ correction
arsing from the interaction of the quasi-2D layer with the unifrom background,
as discussed in subsection~\ref{gau-thin}. Since this depends on the
layer thickness $a$, this correction does not occur in the ideal 2D
system.
\begin{table}
\caption{Results of energy minimization for a hexagonal
  lattice and comparison with the unpolarized
  liquid phase energy $E_L$.
   The energies are measured in $10^{-3}$ atomic
  units.}
\label{tabpol2}
\begin{ruledtabular}
\begin{tabular}{ccccccc}
\hline
$r_s$ \rule{0pt}{2.6ex} &15& $20$ & 30& $50$   \\
\hline 
$E_{cda}$ \rule{0pt}{2.6ex}&-6.7255 & -10.1581&-10.8576  & -9.1112  
 \\
$E_{lda}$ \rule{0pt}{2.6ex}&-6.5036 & -9.8169&-10.7782 & -9.0306 
 \\
$E_{L}$ \rule{0pt}{2.6ex}&-7.1324 & -10.0249 &-10.5995& -8.8939 
\\
\end{tabular}
\end{ruledtabular}
\end{table}
\subsection{Phase transition Liquid$\to$ Wigner crystal.}
\label{wpt-sec}
According to quantum Monte Carlo simulations, the phase transition between a
$\delta$-thin 2D electron liquid  and a 2D electron Wigner crystal occurs
 around $r_s=
35\pm5$.
With our methods we find a transition for $r_s=32$ using the CDA  and
$r_s=38$ using the LDA.

In Figure \ref{figwpt}, we show the phase diagram of the system (in
order to have a clear display we present $(E-E_M){r_s}^{3/2}$ where
$E_M=-1.106103/r_s$ the Madelung energy). The fluid phase energy is calculated
using the fit given by\cite{attac}. Unlike in the $\delta$-thin
2D system, the total energy contains the term $\Delta_{be}=2a/3r_s^2$
arising from the interaction with the unifrom background,
(see subsection~\ref{gau-thin}). This has been removed from both the
liquid and the solid phase 
energies as this improves the clarity of the display.

\begin{figure}
\includegraphics*[width=6cm,height=7.5cm]{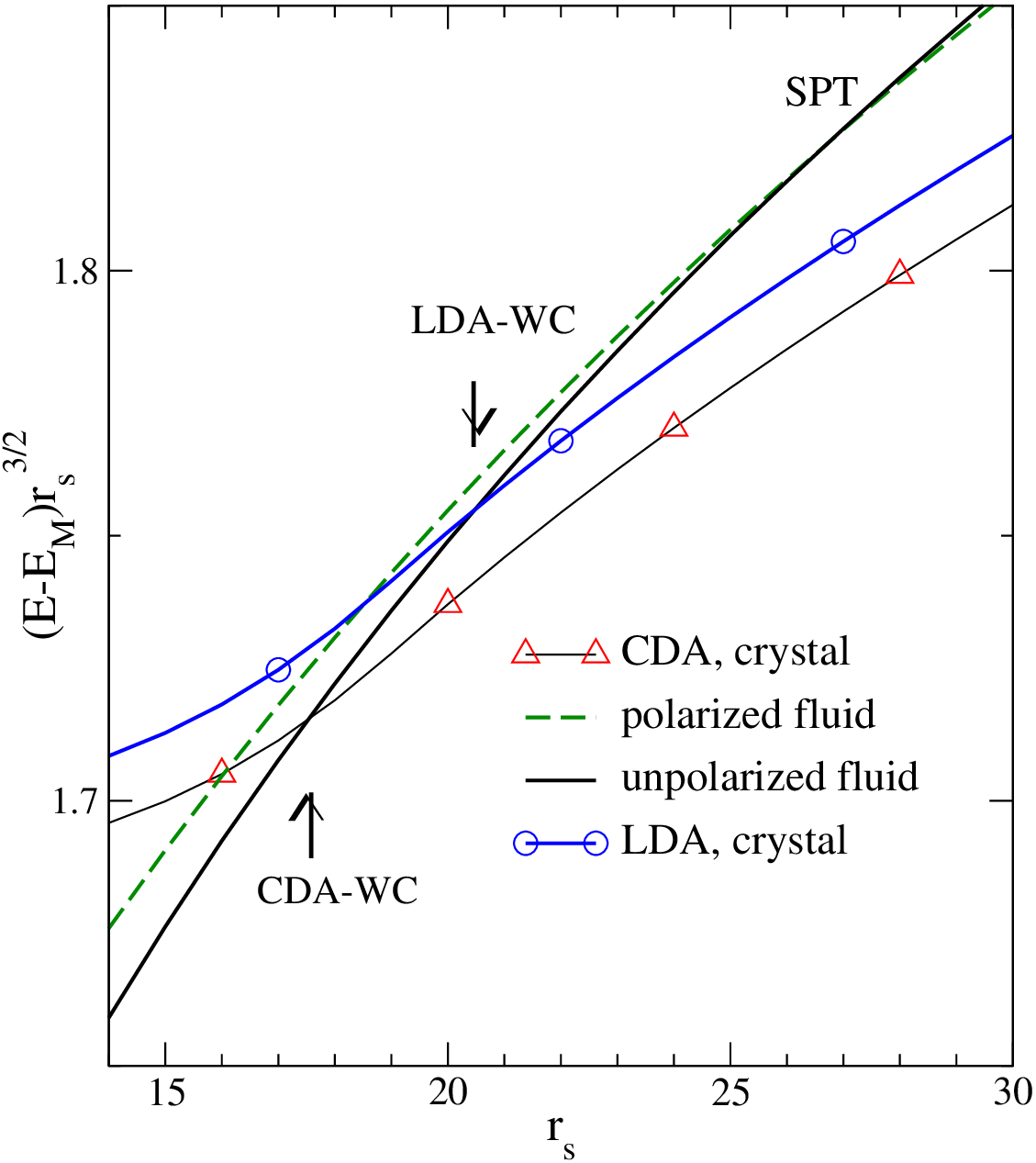}
\caption{(Color online) $(E-E_M){r_s}^{3/2}$ where
$E_M=-1.106103/r_s$ and $E$ is the unpolarized or 
fully-polarized fluid energy, or the solid energy
$E_{cda}$ or  $E_{lda}$. The spin-phase transition in
the fluid is labeled SPT. The onset of the Wigner crystal
in CDA and LDA are indicated by arrows.
}
\label{phaset}
\end{figure}

These results
tend
to show that both LDA and CDA provide an adequate evaluation of  $E_{xc}$ for
 electron solids, especially when we recall that the $E_c$ at $r_s$
values (200-8000),
way outside the fit range ($r_s\le 40$), are needed in the CDA and the
LDA evaluations.
Figure~\ref{phaset} displays the phase transitions
in the quasi-2D HIGFET system. The Wigner transition occurs at
$r_s=15.5$ for the CDA, and $r_s=20.5$ for the LDA, i.e,
before the spin-phase transition (marked SPT in the figure)
of the liquid phase. Since the $\delta$-thin 2D layer is
expected to have a WT near $r_s\sim 35$, the thickness effect
has brought the WT to smaller $r_s$ values. It should be noted
that if correlation effects are neglected, the WT occurs at very low
$r_s$. Hence the shift of the WT to low $r_s$ is a consequence
of the reduced correlations in the quasi-2D system.

\section{Conclusion}

We have investigated the crystallization of electrons under their
own interaction in infinitely-thin 2D electron layers, known
as Wigner crystallization, as well as
in 2D layers with a finite width. The case of
 infinitely-thin 2D electron layers has been extensively studied
in the past, and provides a bench mark to test our methods
for replacing inhomogeneous electron densities by equivalent
uniform-density models. Detailed Hartree-Fock
calculations with large plane-wave basis sets are shown to
be closely equivalent to the single-gaussian harmonic
lattice calculations. We show that the constant-density approximation,
CDA, as well as the local-density approximation, LDA, successfully
recover the total energy as well as the correlation energies of
the infinitely-thin 2D electron crystal. In particular, these
models predict a liquid$\to$solid phase-transition in the
range $30<r_s<40$, in good agreement with Quantum Monte Carlo
simulations. When these methods are applied to quasi-2D
layers with the thickness as in HIGFETS, the weakened 
Coulomb correlations move the Wigner transition towards 
high densities. The LDA and the CDA predict a 
Wigner transition near $r_s\sim$15 to 21.

\appendix*
\section{Evaluation of the electrostatic energy and lattice sums.}
The expression for the electron-electron Coulomb energy, Eq.~\ref{delta-ee}
can be rewritten, using the $\theta$ Jacobi function technique as,
{\setlength\arraycolsep{2pt}
\begin{eqnarray}\label{eq1}
E_{ee}(q) & = &  \frac{N}{2\sqrt{\pi}} \sum_{j}\int_{0}^{\infty}
dy\,y^{-1/2}f(y)e^{-y|\mathbf{\rho_j}|^2-i\mathbf{q\cdot\rho_j}}\nonumber \\
& & \left(\prod_{\alpha} \theta\left [ \frac{a_{\alpha}}{2\pi}
(2\rho_{j}^{\alpha}y+iq_{\alpha});\frac{y {a_{\alpha}}^2}
{\pi}\right] - \delta_{i,0}\right)\nonumber
\end{eqnarray}}
where $\delta_{i,0}$ is the Kronecker symbol,
 because when $\mathbf{\rho_j}=0$, we must have $(m,n)\ne (m',n')$.

\noindent The advantage of introducing the
 Jacobi function $\theta(z,X)$ is that it
  converges well for large $X$ 
and we are also able to obtain convenient
 well-convergent formulas for the small-X
  region by applying the transformation (Eq.(\ref{theta2})) 
from which the Coulomb singular part at $q\rightarrow 0$ 
can be rigorously extracted. Thus, $E_{ee}(q)$ obtained in Eq.(\ref{eq1}) can be 
separated into a large $y$ part and a small $y$ part given by
\begin{equation}
E_{ee}(q)=E_{ee}^{>}(q)+E_{ee}^{<}(q)
\end{equation} 
where
\begin{widetext}
\begin{equation}\label{eee>} 
E_{ee}^{>}(q) = \frac{N}{2\sqrt{\pi}} \sum_{j}
\int_{y_0}^{\infty}dy\,y^{-1/2}
f(y)e^{-y|\mathbf{\rho_j}|^2-i\mathbf{q\cdot\rho_j}}
 \left(\prod_{\alpha=x,y} \theta\left [ \frac{a_{\alpha}}
 {2\pi}(2\rho_{j}^{\alpha}y+iq_{\alpha});
 \frac{y {a_{\alpha}}^2}{\pi}\right] - \delta_{i,0}\right)
\end{equation}
and
{\setlength\arraycolsep{2pt}
\begin{eqnarray}\label{eee<}
E_{ee}^{<}(q) & = & \frac{N}{2\sqrt{\pi}}
 \sum_{j} \int_{0}^{y_0}dy\,y^{-1/2}
f(y)e^{-y|\mathbf{\rho_j}|^2-i\mathbf{q\cdot\rho_j}}
\left(\prod_{\alpha=x,y} \theta\left [ \frac{a_{\alpha}}
{2\pi}(2\rho_{j}^{\alpha}y+iq_{\alpha});\frac{y {a_{\alpha}}^2}{\pi}
\right] - \delta_{i,0}\right) \nonumber \\
& = & \frac{N\sqrt{\pi}}{2 a_1
  a_2}\sum_{j}\int_{0}^{y_0}dy\,y^{-3/2}f(y)e^{-\frac{q^2}{4y}}
  \left(\prod_{\alpha=x,y} \theta\left 
  [ -i\frac{2\rho_{j}^{\alpha}y+iq_{\alpha}}
  {2a_{\alpha}y};\frac{\pi}{y{a_{\alpha}}^2}
  \right]-1+1\right)\nonumber\\
 & & -\frac{N}{2\sqrt{\pi}}\int_{0}^{y_0} dy
 \,y^{-1/2}f(y) \nonumber\\
& = & \frac{N\sqrt{\pi}}{2 a_1
  a_2}\sum_{j}\int_{0}^{y_0}dy\,y^{-3/2}f(y)
  e^{-\frac{q^2}{4y}} \left(\prod_{\alpha=x,y}
   \theta\left [  -i\frac{2\rho_{j}^{\alpha}y+iq_{\alpha}}
   {2a_{\alpha}y};\frac{\pi}{y{a_{\alpha}}^2}\right ]-1\right)\nonumber\\
& & -\frac{n_l N\sqrt{\pi}}{2 a_1
  a_2}\int_{y_0}^{\infty}dy\,y^{-3/2}f(y)
  e^{-\frac{q^2}{4y}} -\frac{N}{2\sqrt{\pi}}\int_{0}^{y_0}
   dy\,y^{-1/2}f(y)+E_{ee}^{hom}(q)
\end{eqnarray}}
\end{widetext}
where $n_l$ is the number of sublattices ($a_1 a_2 /n_l = \pi {r_s}^2$). The
results of the calculation are independent of the value of  $y_0>0$;
nevertheless, we choose it such
that the sums $\theta$ converge fast, and we have
\begin{equation}\label{eeehom}
E_{ee}^{hom}(q)=\frac{n_l N\sqrt{\pi}}{2 a_1  a_2}
\int_{0}^{\infty}dy\,y^{-3/2}f(y)e^{-\frac{q^2}{4y}}
\end{equation}
In order to complete the calculation of $E_{Cou}$, we need to discuss the
form of $\eta(z)$. In their article \cite{fuji2}, Fujiki and Geldart use the
Fang-Howard density $\eta_{fh}(z)=\frac{1}{2 b^3}z^2 e^{-\frac{z}{b}}$ (Figure
\ref{fang}). As already discussed we replace the FH distribution
by the equivalent CDA, i.e., we use a constant density
slab of width $a=16b/3$.
\noindent With this homogeneous form of density 
{\setlength\arraycolsep{2pt}
\begin{eqnarray}
f(y) & = & {\bar{\eta}}^2 \int_{0}^{a} dz \int_{0}^{a} dz'
e^{-y(z-z')^2}\nonumber\\
& = & {\bar{\eta}}^2 \frac{(e^{-a^2 y}+a\sqrt{\pi
    y}\,\mathrm{erf} (a\sqrt{y})-1)}{y}
\end{eqnarray}}
We see here an advantage of the constant density
mapping to density $\bar{\eta}$, the analytic expression
of $f(y)$ being quite simple.

In Eq.(\ref{eed}), we replace $\eta(z')$ by its expression
{\setlength\arraycolsep{2pt}
\begin{eqnarray}\label{eed2}
E_{ed}(q) & = & -\frac{N}{q{r_s}^2}\frac{e^{-q b_d}}{a
  q}(1-e^{-aq})\nonumber\\
E_{ed}(q\rightarrow 0) & = & -\frac{N}{{r_s}^2}
\left(\frac{1}{q}-\frac{a}{2}-b_d+O(q)\right)
\end{eqnarray}}
We recall that $b_d$ is positive or zero, and
gives the location of the donor layer at $z=-b_d$.
Now, in Eq.(\ref{eeehom}), we use the definition of $f(y)$.
{\setlength\arraycolsep{2pt}
\begin{eqnarray}\label{eeehom2}
E_{ee}^{hom}(q) & = & \frac{N}{q{r_s}^2}\frac{2}{q
  a^2}\left(a-\frac{1}{q}(1-e^{-aq})\right)\nonumber\\
E_{ee}^{hom}(q\rightarrow 0) & = & \frac{N}{{r_s}^2}
\left(\frac{1}{q}-\frac{a}{3}+O(q)\right)
\end{eqnarray}}

Now, we will use Eq.(\ref{edd}), (\ref{eed2}),
 (\ref{eee>}), (\ref{eee<}),
(\ref{eeehom2}) in Eq.(\ref{ecoul}) to calculate 
the Coulomb energy for different
types of lattices and for different thicknesses. 
We can see that the expression of $E_{Cou}$ is
dependent on the parameter $b_d$. Since this is a constant
contribution, we set $b_d=0$ and focus on the part
 which depends only on the geometry
of the lattice and on its thickness.

\end{document}